\documentclass[11pt, a4paper, onecolumn]{article}

\usepackage[utf8]{inputenc}
\usepackage[T1]{fontenc}
\usepackage{amsmath, amssymb, bm, braket}
\usepackage{graphicx}
\usepackage{geometry}
\usepackage{titlesec}
\usepackage{fancyhdr}
\usepackage{hyperref}
\usepackage{caption}
\usepackage{subcaption}
\usepackage{appendix}
\usepackage{dcolumn}
\usepackage{enumitem}
\usepackage{array}
\usepackage{diagbox}
\usepackage[numbers, sort&compress]{natbib}
\usepackage{float}
\usepackage{authblk}

\titleformat{\section}{\normalfont\Large\bfseries}{\thesection}{1em}{}
\titleformat{\subsection}{\normalfont\large\bfseries}{\thesubsection}{1em}{}

\begin{document}

\title{\bfseries The Manipulate-and-Observe Attack on Quantum Key Distribution}

\author[1]{William Tighe}
\author[1]{George Brumpton}
\author[2]{Mark Carney}
\author[1]{Benjamin T. H. Varcoe\thanks{Corresponding author: b.varcoe@leeds.ac.uk}}
\affil[1]{School of Physics and Astronomy, University of Leeds, Leeds LS2 9JT, UK}
\affil[2]{Quantum Village Inc., 1007 North Orange Street, Wilmington, DE 19801, USA}

\date{\today}

\maketitle

\begin{abstract}
Quantum key distribution is often regarded as an unconditionally secure method to exchange a secret key by harnessing fundamental aspects of quantum mechanics. Despite the robustness of key exchange, classical post-processing reveals vulnerabilities that an eavesdropper could target. 
In particular, many reconciliation protocols correct errors by comparing the parities of subsets between both parties. 
These communications occur over insecure channels, leaking information that an eavesdropper could exploit. 
Currently there is no holistic threat model that addresses how parity-leakage during reconciliation might be actively manipulated. 
In this paper we introduce a new form of attack, namely the `Manipulate-and-Observe attack' in which  the adversary (1) partially intercepts a fraction $\rho$ of the qubits during key exchange, injecting the maximally tolerated amount of errors up to the 11\% error threshold whilst remaining undetected and (2) probes the maximum amount of parity-leakage during reconciliation, and exploits it using a vectorised, parallel brute force filter to shrink the search space from $2^n$ down to as few as a single candidate, for an $n$-bit reconciled key. We perform simulations of the attack, deploying it on the most widely used protocol, BB84, and the benchmark reconciliation protocol, Cascade. Our simulation results demonstrate that the attack can significantly reduce the security below the theoretical bound and, in the worst case, fully recover the reconciled key material. Our attack presents a proof of concept to target real-world systems where the appropriate hardware is accessible. The main conclusion of this paper is that an integrated security proof of BB84 and Cascade is imperative. The principles of the attack could threaten other parity-based reconciliation schemes, like Low Density Parity Check, which underscores the need for urgent consideration of the combined security of key exchange and post-processing.
\end{abstract}

\section{Introduction}
\pagenumbering{arabic}
\setcounter{page}{1}
Continuous advances in quantum computing and algorithmic development threaten classical cryptographic systems \cite{shorPolynomialTimeAlgorithmsPrime1997, bonehAttackRSAGiven1998, fengSmallPublicExponent2024}; however, Quantum Key Distribution (QKD) offers solutions to these challenges and is claimed to provide information-theoretic security, the pinnacle of cryptography \cite{bennettExperimentalQuantumCryptography1992, PDFQuantumCryptographicNetworks2024, bennettQuantumCryptographyPublic2014, ekertQuantumCryptographyBased1991, niemiecMeasureSecurityQuantum2012, gottesmanProofSecurityQuantum2003, UnconditionalSecurityMemoryBounded, loUnconditionalSecurityQuantum1999, koashiSimpleSecurityProof2009, xuSecureQuantumKey2020, zhangSecurityAnalysisOptimization2020, mehicErrorReconciliationQuantum2020, mehicCalculationKeyLength2015, calverEmpiricalAnalysisCascade2011}. The ideal level of security is `unconditional' where communications are considered to be secure regardless of an eavesdropper's computational capability \cite{xuSecureQuantumKey2020}. In reality, the only way to achieve this is via the `one-time pad' \cite{shannonCommunicationTheorySecrecy1949}, hence  QKD aims to leverage the security of the one-time pad by providing a mechanism for the exchange of large amounts of shared secret key. The idea is to exploit elements of quantum mechanics to guarantee information-theoretic security in the exchange of this key. A number of papers have made significant contributions towards establishing security proofs, and these proofs underpin a vast array of quantum cryptographic systems \cite{bennettQuantumCryptographyPublic2014, ekertQuantumCryptographyBased1991, niemiecMeasureSecurityQuantum2012, gottesmanProofSecurityQuantum2003, UnconditionalSecurityMemoryBounded, loUnconditionalSecurityQuantum1999, koashiSimpleSecurityProof2009, xuSecureQuantumKey2020, shorSimpleProofSecurity2000, brassardSecretKeyReconciliationPublic1994, buttlerFastEfficientError2003, Senekane18, lutkenhausEstimatesPracticalQuantum1999}.

Threat modelling of QKD has also progressed with quantum hacking, drawing attention to previously unconsidered vulnerabilities. Quantum hacking, for example, targets the quantum phase of QKD or exploits equipment weaknesses in practical implementations \cite{xuSecureQuantumKey2020, makarov_FakedStatesAttack2005, vakhitovLargePulseAttack2001a, fungPhaseRemappingAttackPractical2007, qiTimeshiftAttackPractical2006, makarovControllingPassivelyQuenched2009, dusekGeneralizedBeamsplittingAttack1999, amellalQuantumManMiddleAttacks2023, lamas-linaresBreakingQuantumKey2007, khanRobustQuantumCommunication2024}. Examples of attacks include faked state \cite{makarov_FakedStatesAttack2005}, Trojan horse \cite{vakhitovLargePulseAttack2001a}, phase-remapping \cite{fungPhaseRemappingAttackPractical2007}, time shift \cite{qiTimeshiftAttackPractical2006}, and detector blinding \cite{makarovControllingPassivelyQuenched2009}. Another notable attack is photon number splitting (PNS), which uses a hypothetical beamsplitter that can extract a single photon from a multiphoton packet during key exchange \cite{dusekGeneralizedBeamsplittingAttack1999}. PNS has been countered by decoy states and single-source photon emissions \cite{xuSecureQuantumKey2020, lounisSinglePhotonsDemand2000, michlerQuantumDotSinglePhoton2000}. Another strategy is to exploit side channels to initiate an attack. The aim here is to monitor inadvertent leakages from the cryptographic system \cite{parkSingleTraceSidechannel2021}. For example, a power analysis attack exploits the different levels of power used at various stages of an algorithm \cite{kimSingleTraceSide2018}. The development of such attacks instigates efforts to produce countermeasures, leading to better threat modelling and increased security for QKD. 

While most of the effort in securing QKD has been focused on consolidating the quantum phase, there has been a comparative neglect in the development of security in the post-processing phase \cite{parkSingleTraceSidechannel2021, lamas-linaresBreakingQuantumKey2007}. 
During the classical phase, information is conveyed over classical communication channels, which are susceptible to eavesdropping or side-channel attacks \cite{kimSideChannelVulnerability2021}. For this reason, there has been a growing interest in establishing a more robust post-processing threat model \cite{khanRobustQuantumCommunication2024}. Security issues with sifting were explored in \cite{pfisterSiftingAttacksFinitesize2016}, and vulnerabilities in the cache side channels during privacy amplification were investigated in \cite{nikiforovTechnicalReportTUDCS20180024}. A series of side-channel attacks on error reconciliation protocols were also established in \cite{kimSingleTraceSide2018, parkSingleTraceSidechannel2021, kimSideChannelVulnerability2021}, where a single power consumption trace enabled full recovery of the sender's sifted key while operating the Low Density Parity Check (LDPC) error reconciliation protocol \cite{parkSingleTraceSidechannel2021}. This attack has been modified to target any error reconciliation protocol employing parity sums \cite{kimSideChannelVulnerability2021}. 
These attacks stress the importance of threat modelling for post-processing and have prompted countermeasures that harden its security \cite{kimSingleTraceSide2018, parkSingleTraceSidechannel2021, kimSideChannelVulnerability2021}.

Nevertheless, to date, there has been very little effort in understanding whether information leakage in the quantum part of QKD could affect the security of the classical part \cite{khanRobustQuantumCommunication2024}. In this paper, we will explore a novel combined attack on QKD in which we use information leakage from the quantum phase to attack the classical phase. 
For the classical phase, we will use the Cascade Protocol \cite{brassardSecretKeyReconciliationPublic1994}. 
The reason for focusing on Cascade is that, despite its age, it remains the most frequently used protocol and is widely considered the standardised benchmark due to its efficiency and simplicity \cite{zhangSecurityAnalysisOptimization2020, mehicErrorReconciliationQuantum2020, mullerPerformanceCascadeLDPCcodes2024}. Moreover, while there are a number of other protocols in use, there is very little standardisation. 
Notwithstanding this, it will become clear that most of the ideas presented in the current paper can be extended to other protocols.

Although there have been various modifications and optimisations to Cascade, its general mechanisms remain unchanged \cite{martinez-mateoDemystifyingInformationReconciliation2014}. The protocol performs a series of parity comparisons between subsets of both parties' sifted key material, publicly revealing information over insecure classical channels \cite{brassardSecretKeyReconciliationPublic1994}. 
The information leakage has been thoroughly analysed regarding reconciliation efficiency, yet there has been little evaluation of the security aspects \cite{mehicErrorReconciliationQuantum2020, mehicCalculationKeyLength2015, calverEmpiricalAnalysisCascade2011, martinez-mateoDemystifyingInformationReconciliation2014}. 
Security analyses of error reconciliation protocols generally emphasise the number of leaked bits or the parity bits themselves \cite{zhangSecurityAnalysisOptimization2020, buttlerFastEfficientError2003}. 
However, there is more to consider for the true security of Cascade. 
For example an eavesdropper could manipulate aspects of specific protocol {implementation weaknesses} to generate inadvertent information leakage.
This can create conditions in which it becomes possible to run a brute force attack on a reduced reconciled key space. 
Such attacks work, because although the aim of QKD is to simulate a one time pad, the processing of the transmissions and application of error corrections are batched.
The security of the whole therefore comes down to the security of each batch.
QKD in this respect is a rapid rekeying operation, rather than a continuous generation of key material that the one-time-pad application would nominally require.

\begin{figure}[H]
\centering
\includegraphics[width=0.7\linewidth]{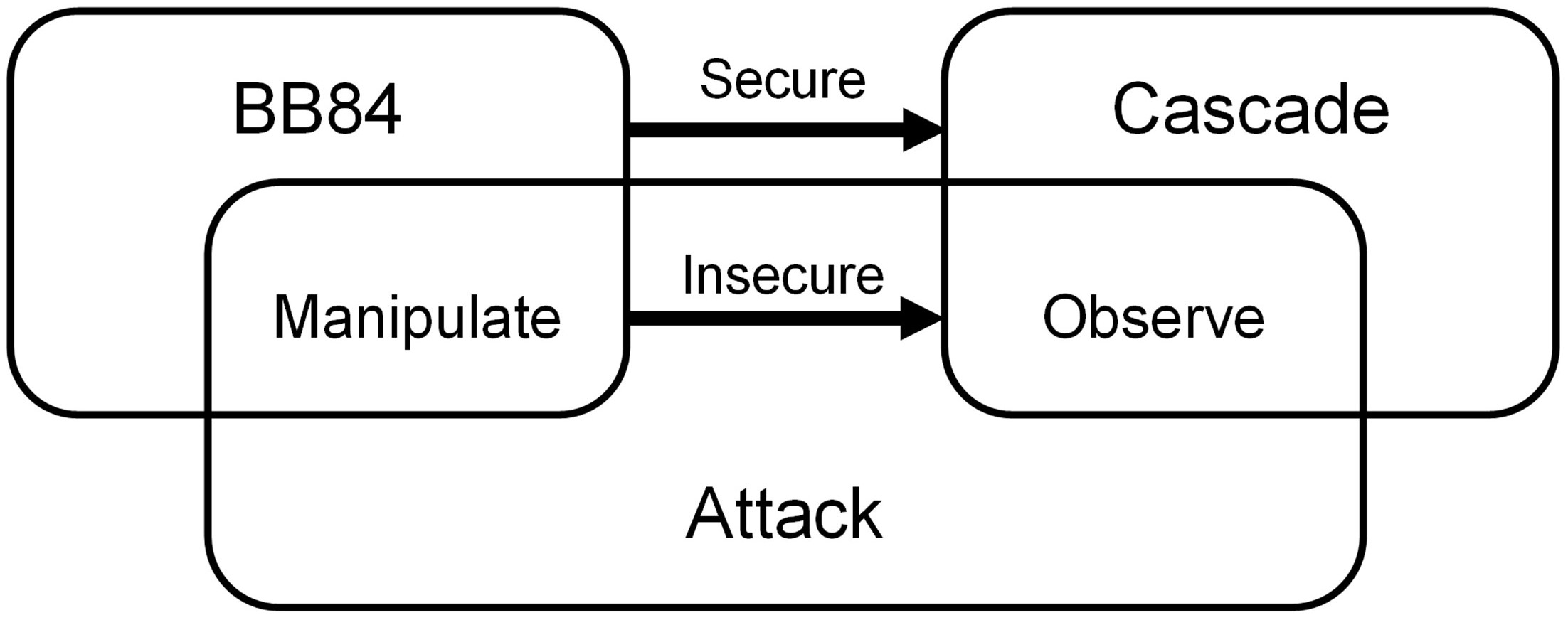}
\caption{\label{Fig1}Threat model diagram of the \textbf{Manipulate-and-Observe Attack}, which exploits the non-extensible security of BB84 and Cascade. For completeness, security proofs of BB84 are given in \cite{bennettQuantumCryptographyPublic2014, gottesmanProofSecurityQuantum2003, loUnconditionalSecurityQuantum1999, koashiSimpleSecurityProof2009, xuSecureQuantumKey2020, shorSimpleProofSecurity2000, Senekane18, lutkenhausEstimatesPracticalQuantum1999} and security analyses of Cascade are given in \cite{mehicErrorReconciliationQuantum2020, calverEmpiricalAnalysisCascade2011, brassardSecretKeyReconciliationPublic1994, mullerPerformanceCascadeLDPCcodes2024, martinez-mateoDemystifyingInformationReconciliation2014, PDFCrackingCurious2024, pedersenHighPerformanceInformation2013}. The attack is split into two phases: the active phase (\textbf{manipulate}) and the passive phase (\textbf{observe}). The active phase consists of a partial intercept-resend attack on BB84, where a fraction of the raw key is `eavesdropped' using an intercept-resend attack. 
The aim here is to obtain some initial knowledge of the raw key material while only injecting up to the maximum tolerated number of errors up to the 11\% error threshold and thereby remaining undetected. 
This has the added advantage of inserting new errors into the raw key material in known places. We can then probe Cascade to leak additional information over public channels, which is observed in the passive phase of the attack. We combine the information gained from both phases of the attack to filter the search space of possible reconciled keys.}
\end{figure}

We have dubbed this attack the `Manipulate-and-Observe Attack'. 
The root of this attack arises because security is not extensible across a number of separate components, as depicted in Figure~\ref{Fig1}. We employ active eavesdropping strategies during the quantum phase to manipulate post-processing vulnerabilities. A weakness in the error estimation protocol is exploited with a partial intercept-resend attack, which allows us to obtain a fraction of the raw key material whilst remaining undetected. Simultaneously, the maximally tolerated number of errors is injected into the receiver's raw key, probing Cascade to leak the maximum amount of parity information. 
We apply a vectorised, parallel brute force filter to reduce the search space from $2^n$ to as few as a single candidate for an $n$-bit reconciled key. 
Through simulations, we oberve that it is possible to reduce the security significantly below the theoretical bound and we have observed circumstances in which we can obtain a full recovery of the reconciled key in the worst case. 
Therefore, we demonstrate that the combined security of BB84 and Cascade is not information-theoretic but computational, and a holistic view of the combined system's security must be considered. 
The attack could be modified to target other error reconciliation protocols that utilise parity checks, like LDPC. 
Due to the vast use of such protocols, the attack could seriously threaten the rapid adoption of QKD networks \cite{xuSecureQuantumKey2020}.

From an information theoretic point of view, the issue exploited in this work arises from a fundamental shift from passive to active attacks from Eve. As a passive observer, the general security assumption relies on the mutual information between Alice, Eve, and Bob being $$ I(E|A;B) < I(A;B) $$ which for Eve being a \emph{passive} observer is correct. However, if Eve is an \emph{active} participant in the manner we describe in this paper, then the retransmission from Eve to Bob after a measurement on Alice shifts Alice and Eve to being a first-order Markov source \cite{Ash1990}; Eve's manipulations are based on immediately available, if approximate, data. That the outcomes from this re-transmission are unbalanced in favour of Eve is what enables the significance of this attack.

It is important to emphasise that the vulnerabilities explored here arise prior to privacy amplification. Although privacy amplification is frequently described as a mechanism that `restores' security, in practice it cannot undo structural information leakage. It simply compresses the post‑reconciliation key to the minimum length compatible with the adversary’s remaining uncertainty. Our attack reduces this uncertainty directly by shrinking the post‑Cascade keyspace itself, thereby lowering the secure minimum that privacy amplification can produce. In other words, privacy amplification does not mitigate the weakness exploited here; rather, the weakness reduces the effectiveness of privacy amplification.

More broadly, QKD is often described as offering `unconditional' or `physics‑based' security. However, in finite‑size regimes all guarantees are intrinsically statistical. The fact that the security of BB84+Cascade can be compromised by statistically unfavourable and plausible fluctuations in QBER estimation reveals that the practical security of QKD is not solely determined by physical principles, but also by sampling randomness. At the scale of trillions of keys generated per day, statistically rare events become operationally relevant. This motivates treating QKD’s security as statistical rather than absolute, and strengthens the case for analysing adversarial strategies that exploit such deviations.

Although this work focuses on Cascade, the underlying mechanism we exploit, namely structured leakage of parity information during reconciliation, is shared across all parity‑based protocols, including widely‑used LDPC codes. Cascade is chosen here not because it is uniquely vulnerable, but because it is one of the few reconciliation protocols with an established analytical framework and partial security proofs. The attack we present therefore targets the class of parity‑based reconciliation schemes generally, not merely one historical implementation.

Throughout this paper, we distinguish  between the various stages of key material generated in a QKD session. Our attack operates on the key material prior to privacy amplification. Specifically, we are attacking the sifted and reconciled key material produced after the combination of BB84 and Cascade. We therefore use the term `key material' to refer to pre‑Privacy Amplified data. The term `secret key' is reserved exclusively for the final, privacy‑amplified key as defined in composable security proofs. Our results show that the Manipulate‑and‑Observe attack can substantially reduce the entropy of the reconciled key material, thereby reducing the secure minimum available to privacy amplification. Privacy amplification does not restore security here; it can only compress whatever entropy remains. Thus, our attack targets the integrity of the pre‑key upon which the final key’s security ultimately depends.

The outline of this paper is as follows: Section 2 describes Cascade's information leakage, which is exploited to filter a set of reconciled keys, forming the passive phase of the Manipulate-and-Observe Attack. The active phase of the attack is then presented, which employs partial eavesdropping during BB84's quantum key exchange. In particular, we describe the partial intercept-resend attack and how it can be used to manipulate weaknesses in post-processing. Our main results are presented in Section 3, where we simulate the full attack on BB84 and Cascade. In Section 4, we conclude the attack's performance and suggest further improvements.

\section{Manipulate-and-Observe Attack}
\subsection{Cascade}
Cascade \cite{brassardSecretKeyReconciliationPublic1994} is an interactive reconciliation protocol, where errors are corrected through two-way communications between Alice and Bob \cite{calverEmpiricalAnalysisCascade2011}. The protocol is often integrated within the post-processing phase of BB84 \cite{martinez-mateoDemystifyingInformationReconciliation2014}. Detailed descriptions of BB84 \cite{bennettQuantumCryptographyPublic2014} are given in \cite{xuSecureQuantumKey2020, Senekane18} and we have operated the code as if the protocol is being performed under the normal operating conditions. 
We will use the standard nomenclature where the sender, receiver, and eavesdropper are commonly referred to as Alice, Bob, and Eve respectively. 
Discrepancies between Alice's and Bob's sifted keys are referred to as the quantum bit error rate (QBER) \cite{xuSecureQuantumKey2020}. In this article, we follow the recommendation to use a sample size of 37\% of the sifted key material during QBER estimation, as suggested for advanced security \cite{niemiecMeasureSecurityQuantum2012}. Additionally, we set the lower bound for the threshold QBER to 11\%, as suggested in \cite{gottesmanProofSecurityQuantum2003, zhangSecurityAnalysisOptimization2020}. 

\begin{figure}[H]
\centering
\includegraphics[width=1\linewidth]{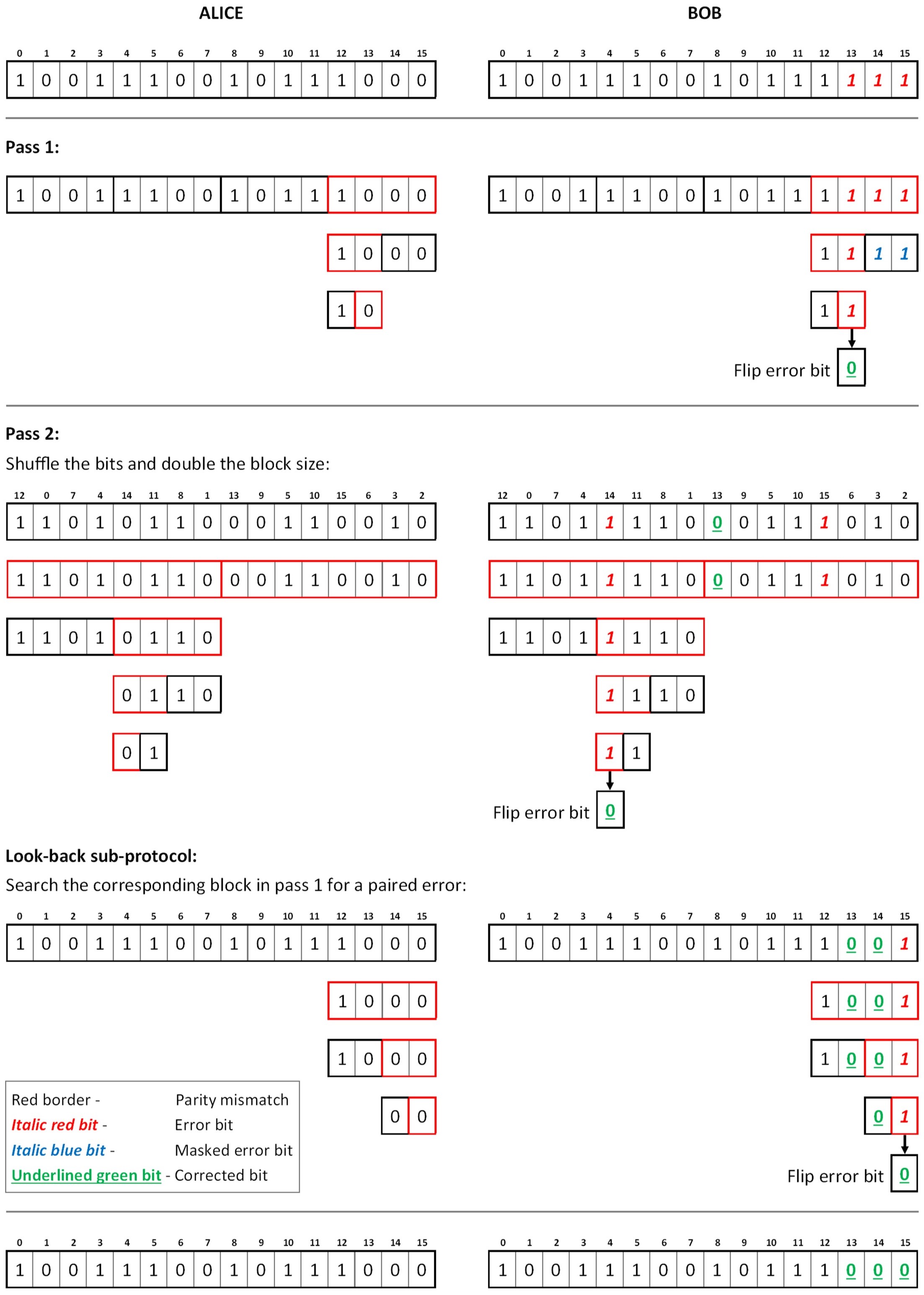}
\caption{\label{Fig3} Diagram of Cascade using two passes to reconcile three errors. During Pass 1, Alice and Bob split their sifted keys into blocks and compare parities. In this example, the parity mismatches in the final block. A search algorithm (`Binary') is run on the final block to locate and correct an error. The parities of the first half of the blocks are compared, revealing another mismatch. The previous step is repeated until the error is located and corrected. During Pass 2, the bits are shuffled to distribute the errors throughout the sifted key material evenly. The block size is doubled, and Binary is applied to the first block due to a parity mismatch. For every pass i>1, whenever an error is corrected, a look-back sub-protocol is activated to correct paired errors that were masked in previous passes. In this example, look-back finds a paired error in the first pass.}
\end{figure}

\clearpage

A consequence of Cascade's public parity comparisons is that information about the reconciled key is leaked. 
The maximum number of leaked bits for checking a block containing an error is $1 + \lceil \log_2(k_i)\rceil$, where the log gives the number of divisions of the block during Binary \cite{mehicCalculationKeyLength2015}. The maximum number of leaked bits throughout all of Cascade is given by \cite{mehicCalculationKeyLength2015}:

\begin{equation}
    L_{max} = \sum_i \left(\frac{n}{k_i} + \sum_{\substack{\text{errors} \\ \text{corrected}}}\lceil \log_2(k_i)\rceil \right)
    \label{eq:3.1}
\end{equation}

\noindent where block size is $k_i = 2k_{i-1}$, $k_i<\frac{n}{2}$ and the $n$ is the length of the sifted key prior to reconciliation \cite{mehicCalculationKeyLength2015}. Hence, the initial block size, thus the QBER, influences the number of leaked bits.

Conversely, the minimum amount of leaked information for reconciliation is given by the conditional entropy \cite{mullerPerformanceCascadeLDPCcodes2024}:
\begin{equation}
    L_{min} = n\mathit{H}(A|B)
        \label{eq:3.2}
\end{equation}

\noindent where we assume that Alice and Bob communicate via a binary symmetric channel $BSC(p)$, and each bit has a probability $p$ of being exposed to noise \cite{brassardSecretKeyReconciliationPublic1994}. The conditional entropy of a $BSC(p)$ is given by \cite{mullerPerformanceCascadeLDPCcodes2024}:

\begin{equation}
    \mathit{H}(A|B) = \mathit{h}(p) = -p\log_2(p) - \left(1-p\right)\log_2(1-p)
\end{equation}

\noindent In the original description of Cascade, the information per block is bounded by \cite{brassardSecretKeyReconciliationPublic1994}:

\begin{equation}
    I(\omega) \le 2 + \frac{1-(1-2p)^{k_1}}{2}\lceil \log(k_1)\rceil + \sum_{l=2}^{\omega}\frac{1}{2^{l-1}}\left(k_1p - \frac{1-(1-2p)^{k_1}}{2} \right)\lceil \log(k_1) \rceil
    \label{eq:3.4}
\end{equation}

\noindent where $\omega$ is the number of passes. 

Equations~\ref{eq:3.1}, \ref{eq:3.2} and \ref{eq:3.4} indicate that the information leakage increases as the QBER increases. Therefore, an eavesdropper should inject the maximum QBER (below BB84's 11\% threshold) into Bob's raw key material to allow Cascade to leak the most information. 

\subsection{The Passive Phase}
The reconciled key material shared between Alice and Bob consists of an array of binary bits. If an eavesdropper has no information about the reconciled key material, there are $2^n$ possible reconciled key candidates to check for a reconciled key size of $n$ bits. 
Each bit revealed to an eavesdropper halves the number of possible candidates immediately after reconciliation. The possible candidates are constrained particularly by the parity conditions leaked by error reconciliation. A parity condition is a tuple of the bit indices in a specific block/sub-block and its parity. The candidates that meet all the parity conditions immediately after reconciliation and before privacy amplification are referred to as 'valid keys'. We define the search space $S$ as the number of valid keys. Hence, the initial search space before any Cascade passes is $S_{\text{int}} = 2^n$.

\begin{figure}
\centering
\includegraphics[width=0.93\linewidth]{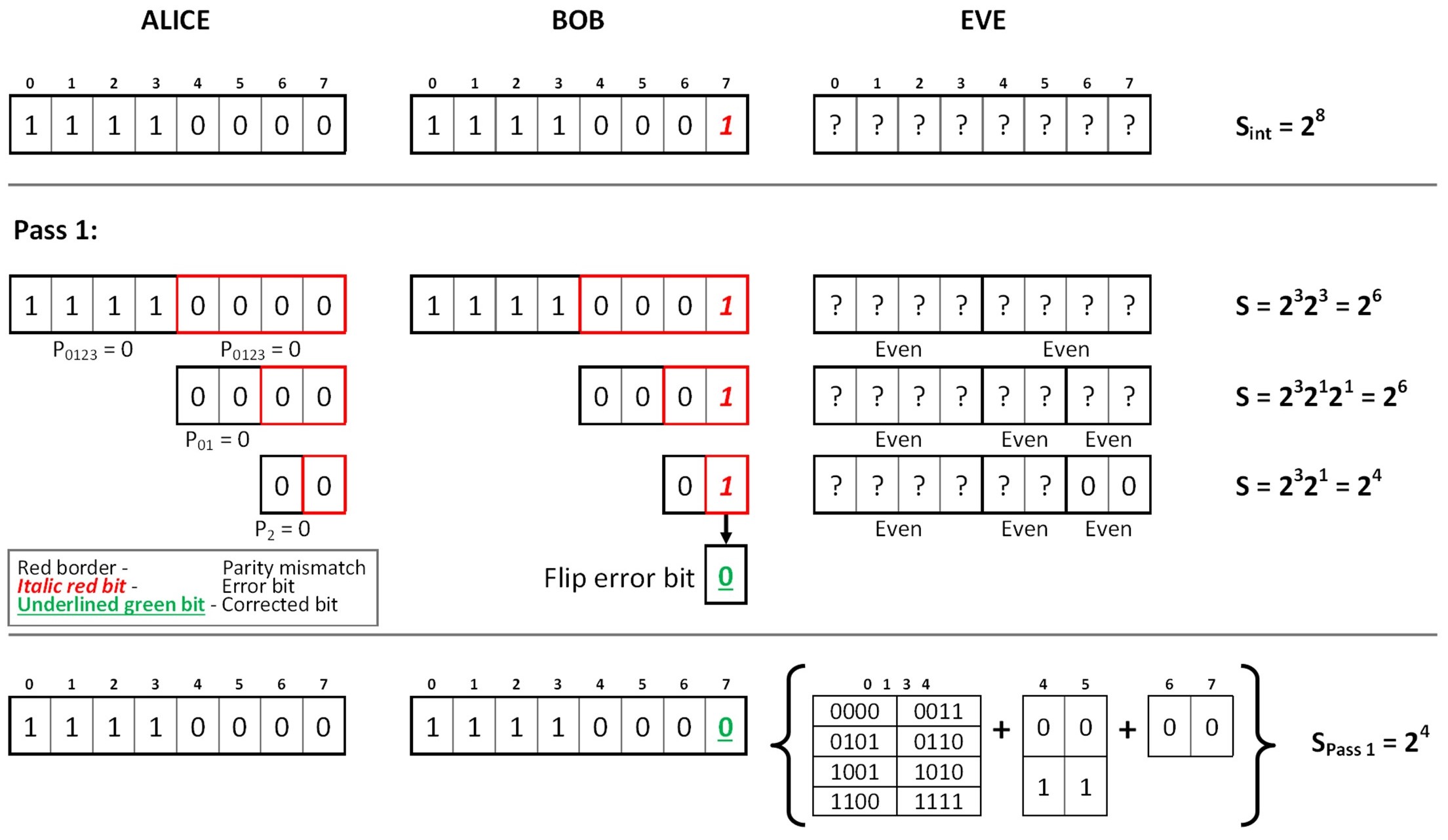}
\caption{\label{Fig4} Diagram of brute force reconciled key filtering using the parity conditions leaked by a single Cascade pass. The parities of the first block are compared, revealing that Alice’s first block has a parity of 0 and must contain an even number of 1s. As the parities of the first block match, Cascade continues to the second block, where the process repeats. This time, the parity of Alice’s first block is 1, so the block must have an odd number of 1s. Alice’s and Bob’s parities for the second block disagree, so Binary is executed, revealing additional parities of the sub-blocks until Alice and Bob locate and correct the error. Each parity check made by Alice and Bob during this first pass halves Eve’s search space.}
\end{figure}

The leaked information can be exploited by Eve to reverse engineer a set of valid keys. The simplest way to determine the valid keys and Eve's search space is to filter all the possible reconciled keys by brute force. For example, one could filter all the possible blocks by their corresponding parity conditions and combine them to get all the valid keys. Figure~\ref{Fig4} demonstrates a simple example of this approach, for an 8-bit pre-reconciled key with a single pass of Cascade and a block size of 4 bits. In this example, Eve can deduce that there are eight valid options for the first block and two valid options for the second block. Hence, her overall search space is $S=2^{3}2^{1}=2^4=16$. We notice that after the first pass, the search space is given by:

\begin{equation}
    S_{\text{pass 1}} = 2^{n-l}
\end{equation}

\noindent where $n$ is the number of bits in the reconciled key and $l$ is the number of leaked bits.

\begin{figure}[H]
\centering
\includegraphics[width=1\linewidth]{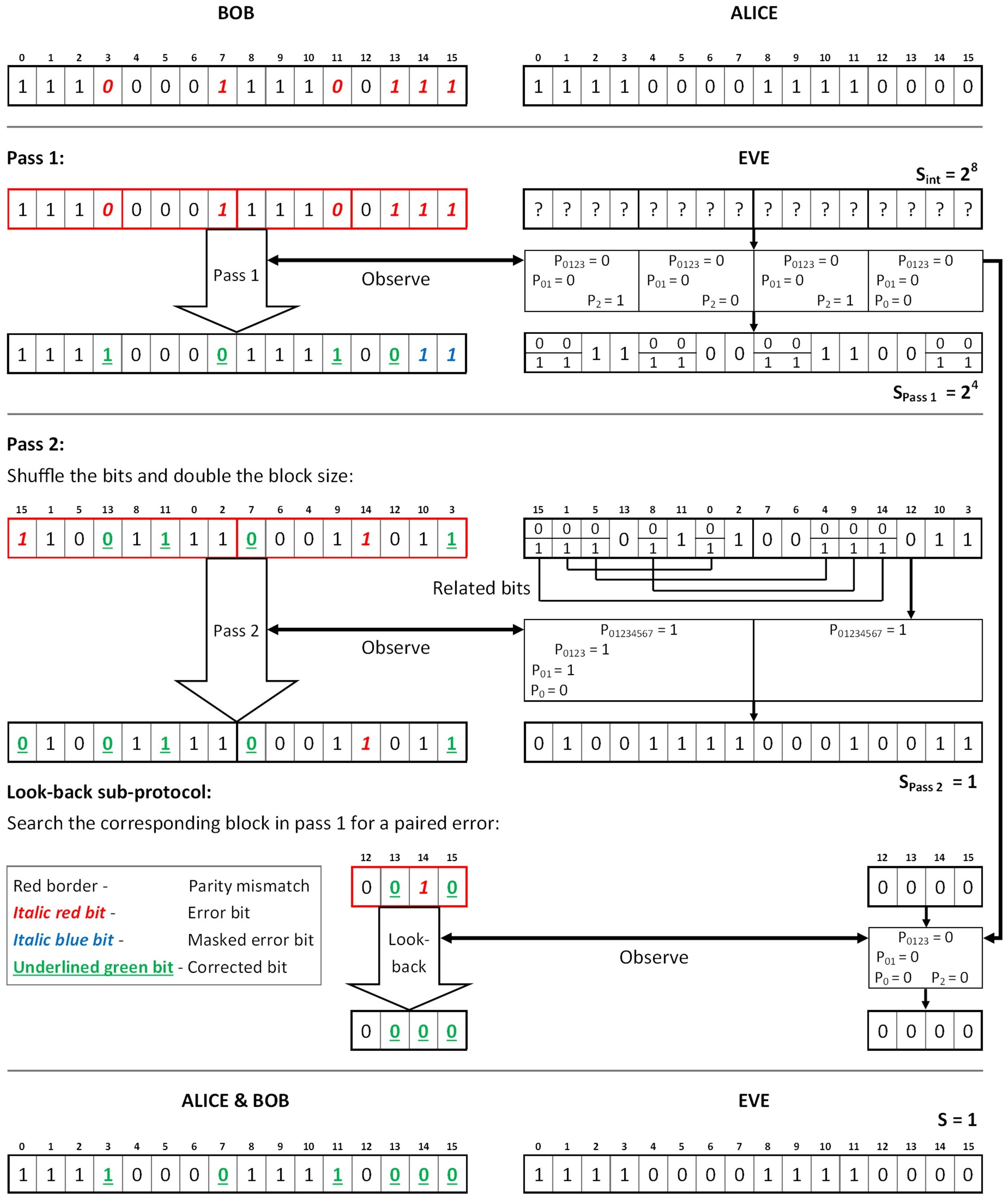}
\caption{\label{Fig5} Diagram of brute force reconciled key filtering using the parity conditions leaked by two Cascade passes. During the first pass, Binary is applied to correct a single error in each block, revealing parity information. Two errors are masked in the final block. Eve observes the leaked parity information and narrows down her search space to $S_{\text{pass 1}}=2^{4}$. During the second pass, the bits are shuffled, and in this case, the masked errors (bits 14 and 15) are distributed across different blocks. Binary is triggered on the first block, corrects bit 14 and leaks additional information. As before, Eve exploits the leaked information to reduce her search space. When Eve uncovers bits in pass 2, it reveals bits in other locations, as these bit pairs are constrained by the parity conditions leaked in pass 1. In this example, Eve has reduced her search space to a single candidate after pass 2, and the information leaked by the look-back sub-protocol is redundant.}
\end{figure}

However, multiple passes of Cascade are more complicated. Now consider a 16-bit pre-reconciled key with 4-bit blocks, where the first three blocks contain a single error, and the final block contains three errors, illustrated in Figure~\ref{Fig5}. In this example, two masked errors are shuffled to different blocks in the second pass, triggering Binary and allowing Eve to uncover the entire reconciled key. However, if the two masked errors were shuffled to the same block, they would remain masked for the second pass. In this case, the protocol would fail to correct the errors and leak less information. Thus, Eve would face greater ambiguity and a larger search space. Simulation results in appendix B indicate that random shuffling provides an additional constraint layer, thereby varying Eve's search space. Sometimes, the parity checks after shuffling are redundant and do not reduce the search space, whereas at other times they significantly reduce it. The simulation results indicate that the search space after all Cascade passes is given by:

\begin{equation}
    S_{\text{Cascade}} = 2^{n-u}
\end{equation}

\noindent where $u$ is the number of unique, non-redundant parity checks. 

\begin{figure}[H]
\centering
\includegraphics[width=0.63\linewidth]{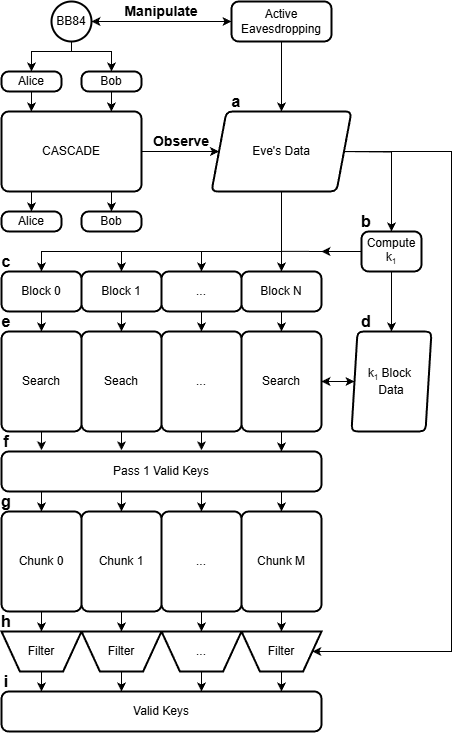}
\caption{\label{Fig6} Diagram of the passive phase of the \textbf{Manipulate-and-Observe Attack}. a) Eve stores the leaked information from Cascade/Binary. b) Eve determines the initial block size $k_1$ and assesses whether a remainder block is required. c) Eve divides her initial key prior to reconciliation into blocks of size $k_1$. d) Eve loads an array that contains all the possible $k_1$ bit blocks and their corresponding parity blocks. e) In parallel, an array of valid blocks is determined for each block index in pass 1 using the vectorised block search. f) Each array of valid blocks is combined to form an array of valid keys for pass 1. g) The array of pass 1 valid keys is split into chunks. h) Brute force is used to filter the pass 1 valid keys in each chunk in parallel. i) The valid keys in each processed chunk are recombined. The remaining valid keys satisfy all constraints imposed by Cascade (and any prior eavesdropping results). If there is one valid key remaining, then Eve has obtained Alice’s and Bob’s reconciled key.}
\end{figure}

A brute force reconciled key filtering approach is simple but inefficient, as every possible block must be filtered through the collection of parity checks. We address this issue with an optimised method that employs a vectorised, parallel filter, which is detailed in appendix C. The optimised method forms the passive phase of the Manipulate-and-Observe Attack, as illustrated in Figure~\ref{Fig6}.

\subsection{The Active Phase}
\begin{figure}[h]
\centering
\includegraphics[width=0.95\linewidth]{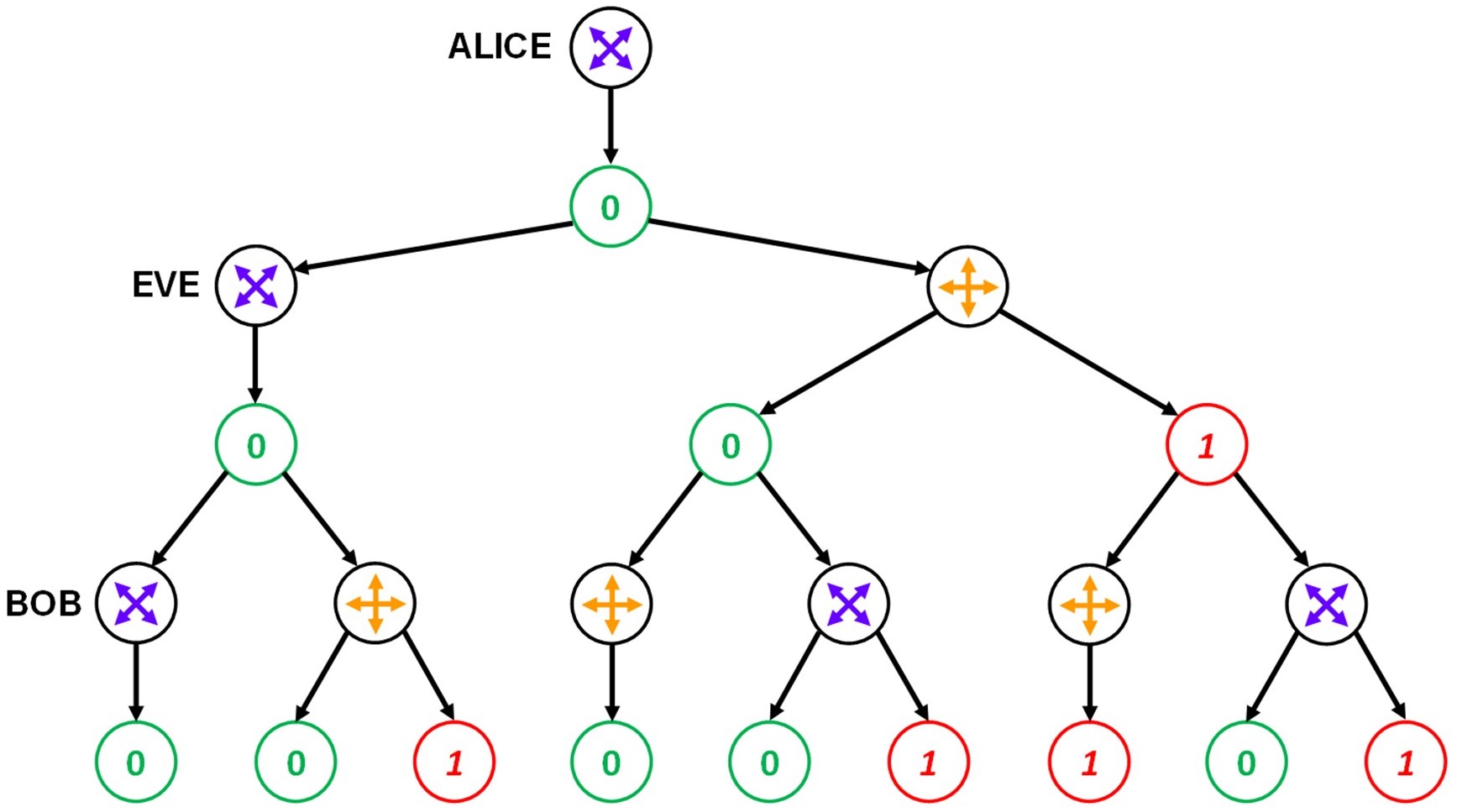}
\caption{\label{Fig7} Tree diagram of an intercept-resend attack on the BB84 protocol. Eve intercepts the qubits during their transmission through the quantum channel and measures using a random basis choice \cite{xuSecureQuantumKey2020}. Next, Eve uses the same basis for re-encoding before sending the new qubit to Bob \cite{xuSecureQuantumKey2020}. When Eve selects the same basis as Alice, she will decode the qubit correctly and introduce no error for Bob to detect. Conversely, if Eve's and Alice's bases differ, there is a 50\% chance that Eve decodes the wrong bit value. Moreover, if Bob uses the same basis as Alice, he has a 50\% chance of decoding the wrong bit value. Thus, Bob will have a QBER of 25\% after sifting. Eve can sift her raw key similarly to Alice and Bob, where she discards the bits where she disagrees with either. She can have complete confidence in her remaining bit values, assuming an ideal system without noise.}
\end{figure}

Despite the optimisations made to the passive phase of the attack, filtering large reconciled keys remains computationally infeasible. We can overcome this computational demand with active eavesdropping strategies targeting QKD's quantum phase. The most basic eavesdropping strategy is the intercept-resend attack, illustrated in Figure~\ref{Fig7} \cite{bennettExperimentalQuantumCryptography1992}. 

A partial intercept-resend attack can allow Eve to reduce her presence by eavesdropping on a subset of the qubits \cite{fioriniEstimatingInterceptionDensity2024}. In this type of attack, the percentage of qubits that Eve intercepts are given by the interception density parameter, $\rho$ \cite{fioriniEstimatingInterceptionDensity2024}. Hence, $\rho$ is the probability that Eve intercepts an individual qubit. The partial intercept-resend attack instigates a QBER given by \cite{fioriniEstimatingInterceptionDensity2024}:  

\begin{equation}
    p = \frac{\rho}{4}
\end{equation}

Therefore, if Eve intercepts 40\% of the qubits, then she will introduce a QBER of 10\% on average, which is less than the maximally tolerated QBER of 11\%. Thus, relying on the assumption that Eve measures the entire raw key material is na\"ive since partial eavesdropping attacks can yield a significant fraction of the raw key whilst remaining undetected.

Statistical fluctuations could favour eavesdropping attempts. For instance, the QBER could be underestimated by comparing a sample of the sifted keys that exclude enough errors. If the estimated QBER were below the threshold, the protocol would continue with Eve undetected. It is claimed that systems which require the highest level of security should compare at least 37\% of the sifted keys during QBER estimation \cite{niemiecMeasureSecurityQuantum2012}. Despite recommending 37\%, results show a noticeable error in the QBER estimation \cite{niemiecMeasureSecurityQuantum2012}. Other works take samples of 10\% of the sifted keys \cite{zhangSecurityAnalysisOptimization2020}. The corresponding error and variation in the QBER estimation for samples of this size are much more prominent \cite{niemiecMeasureSecurityQuantum2012}.

\begin{figure}[h]
\centering
\includegraphics[width=0.75\linewidth]{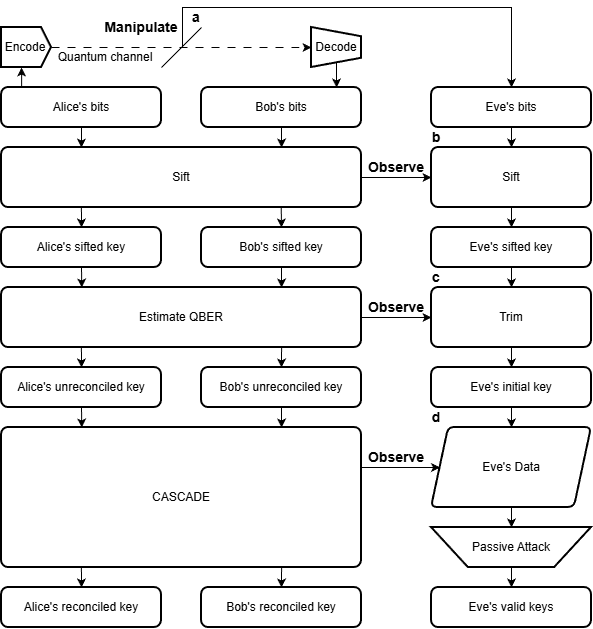}
\caption{\label{Fig8} Diagram of the active phase of the \textbf{Manipulate-and-Observe Attack}. a) Eve partially eavesdrops during the quantum key exchange, obtaining some bits and introducing errors into Bob’s data. b) Eve exploits Alice’s and Bob’s public comparison of their basis choices during sifting, allowing her to sift her raw key. c) Eve discards the corresponding bits Alice and Bob used to estimate the QBER. If the QBER is below the threshold, then Cascade is used to reconcile the sifted keys, leaking information. d) Eve executes the passive phase of the attack that generates a set of valid keys that satisfy all leaked information after reconciliation.}
\end{figure}

We can exploit this vulnerability with partial eavesdropping. Eve's objective is to intercept a fraction of the raw key material such that the QBER she introduces is underestimated by Alice's and Bob's inadequate sample size. Additionally, Eve intends to maximise the error rate whilst keeping it below the threshold. The aim is to manipulate the subsequent Cascade implementation so that it leaks a maximum amount of information. Eve can then use the passive phase of the attack to further reduce her search space. The full Manipulate-and-Observe Attack is depicted in Figure~\ref{Fig8}.

\clearpage

\section{Results and Discussion}
Simulations of the attack were implemented in Python and ran on an AMD Ryzen 5 2600 CPU @3.4GHz, with 16GB of RAM. All 6 CPU cores were utilized in parallel whilst filtering the pass 1 blocks and chunks for the subsequent passes. Due to hardware limitations, we restricted the reconciled key size to $n=100$ bits. Particularly, the memory demand rapidly grows as the number of pass 1 valid keys to store increases with increasing reconciled key size. We consider this reconciled key size sufficient to show a proof of concept of the attack, which could be further scaled across a GPU cluster and run with a low-level language like C++. Furthermore, only three passes of Cascade are simulated to keep the reconciled key size manageable. Since the final Cascade pass corrects a negligible number of errors, we consider three passes to be a suitable approximation \cite{mehicCalculationKeyLength2015, mehicErrorReconciliationQuantum2020}.

Random raw keys and eavesdropping indices were generated using a shared seed, unique for each independent simulation. Likewise, three unique seeds were used to permute the bits for each Cascade pass. Simulations of partial-intercept resend attacks with various eavesdropping rates were performed to obtain a set of seeds where Eve was undetected. We define successful seeds as the seeds where the QBER is underestimated below the threshold, leaving Eve unnoticed by Alice and Bob. Due to BB84's random processes, the successful seeds lead to various reconciled key sizes. To keep the reconciled key size constant, the successful seeds were filtered to obtain a set of reconciled keys with 100 bits. The full attack was deployed exhaustively on these final seeds until 20 results were obtained for each value of QBER. BB84 and Cascade have been simulated with 4.28 million raw keys to obtain these results.

Details of the simulation parameters are provided in Table~\ref{Table 2}. The raw key size was optimised to maximise the number of 100-bit reconciled keys. For instance, an raw key size of 317 bits would be trimmed by approximately 50\% after the transmission of the qubits, as the basis choice bias delta was set to 0.5 like in \cite{zhangSecurityAnalysisOptimization2020}. After sifting, the remaining bits are further trimmed after taking a sample of 37\% for QBER estimation, as recommended for advanced security \cite{niemiecMeasureSecurityQuantum2012}. If the QBER is below the threshold of 11\%, as in \cite{zhangSecurityAnalysisOptimization2020}, the sifted key is further trimmed to a size of around 100 bits. Simulations of the QKD protocol terminate after Cascade since these works primarily focus on the vulnerability of reconciliation. Hence, further sub-protocols like privacy amplification are not considered. Additionally, we consider an idealised system with a perfect quantum key exchange, where no errors stem from equipment imperfections. We acknowledge that real-world implementations of BB84 are subject to noise, which may reduce the proportion of the raw key Eve can obtain through partial eavesdropping whilst keeping the QBER below the allowed threshold. However, we consider an ideal system sufficient to demonstrate a proof of concept for the attack. We provide additional details of the results in Appendix E.

\begin{table}[h]
    \centering
    \begin{tabular}{l l}
        \hline
        \textbf{Property} & \textbf{Value} \\
        \hline
        Raw key size & 317 bits \\
        Reconciled key size & 100 bits \\
        Basis choice bias delta & 0.5 \\
        Eve basis choice bias delta & 0.5 \\
        Detector noise rate & 0 \\
        QBER estimation sampling rate & 0.37 \\
        QBER threshold & 0.11 \\
        Cascade passes & 3 \\
        \hline
    \end{tabular}
    \caption{\label{Table 2}Simulation parameters for partial intercept-resend and Manipulate-and-Observe attacks.}
\end{table}

\subsection{Partial Intercept-Resend Results}
The following results show the performance of the partial intercept-resend attack for different eavesdropping rates $\rho$. The leaked information per block for three Cascade passes, $I(3)$, was computed using Equation~\ref{eq:3.4}. The theoretical bound for Cascade's security was determined by subtracting the product of $I(3)$ and the number of pass 1 blocks $m$, from the reconciled key size $n$. A similar approach was used to calculate the maximum security after reconciliation, where Equation~\ref{eq:3.2} was used to determine the minimum information leakage. After reconciliation, the number of secure bits following the partial intercept-resend attacks is determined by shifting the proportion of the raw key material obtained below Cascade's theoretical bound.

\begin{figure}[h]
\centering
\includegraphics[trim=0 0 0 25, clip, width=0.75\linewidth]{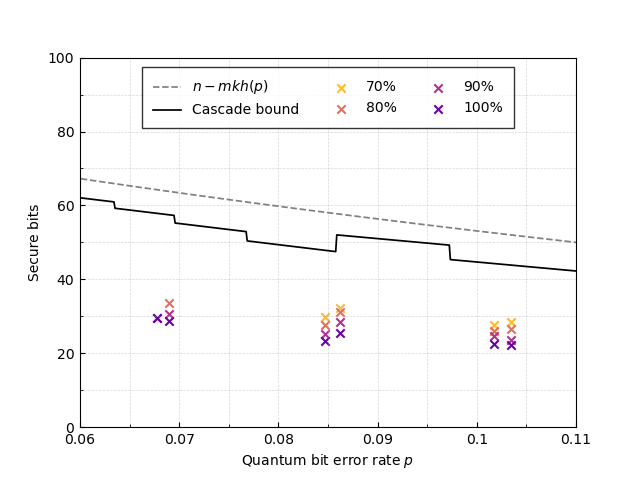}
\caption{\label{Fig10} The number of secure bits after reconciliation, using partial intercept-resend attacks with different eavesdropping rates $\rho$, across various levels of estimated QBER $p$. The security is compared with Cascade's bound and the theoretical maximum security after reconciliation.}
\end{figure}

Figure~\ref{Fig10} demonstrates that a partial-intercept resend attack can yield a large proportion of the key material. Specifically, around 22-33 bits are secure following the attack for all $p$ and $\rho$. As $\rho$ increases, Eve obtains a more significant proportion of the raw key, but the frequency of successful attacks decreases, as shown in Appendix E. Thus, Eve must consider a balance between the performance of the attack and the probability of succeeding. Note that for a QBER of 0.0862, the security briefly increases as the Cascade's bound increases at the same point, due to better block-size selection.

\subsection{Manipulate-and-Observe Results}
\begin{figure}[h]
    \centering
    \begin{subfigure}[b]{0.49\linewidth}
        \includegraphics[trim=20 8 40 25, clip, width=\linewidth]{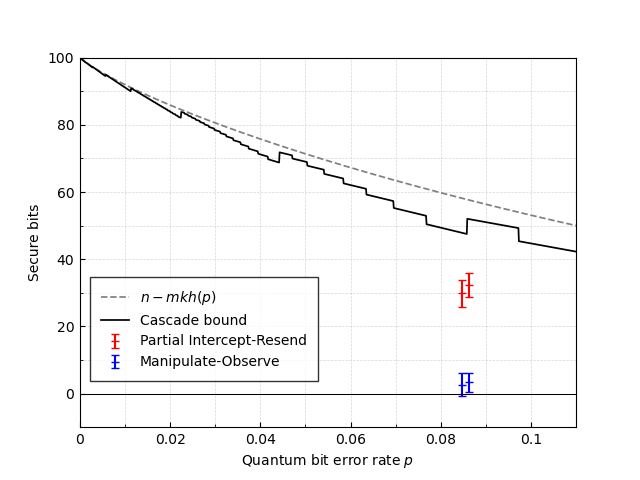}
        \subcaption{$\rho = 0.7$}
    \end{subfigure}
    \hfill
    \begin{subfigure}[b]{0.49\linewidth}
        \includegraphics[trim=20 8 40 25, clip, width=\linewidth]{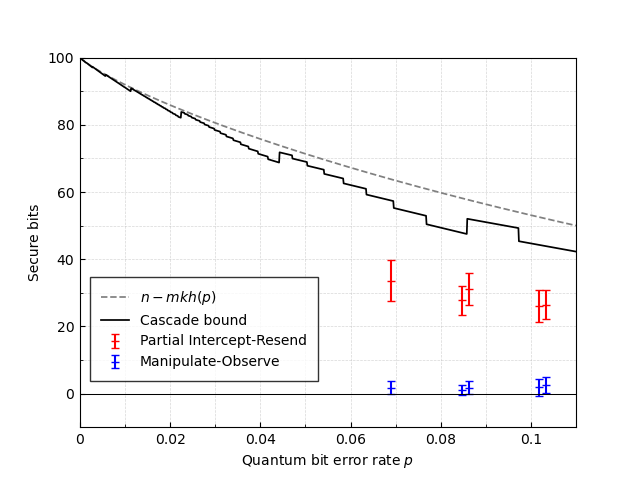}
        \subcaption{$\rho = 0.8$}
    \end{subfigure}
    \begin{subfigure}[b]{0.49\linewidth}
        \includegraphics[trim=20 8 40 25, clip, width=\linewidth]{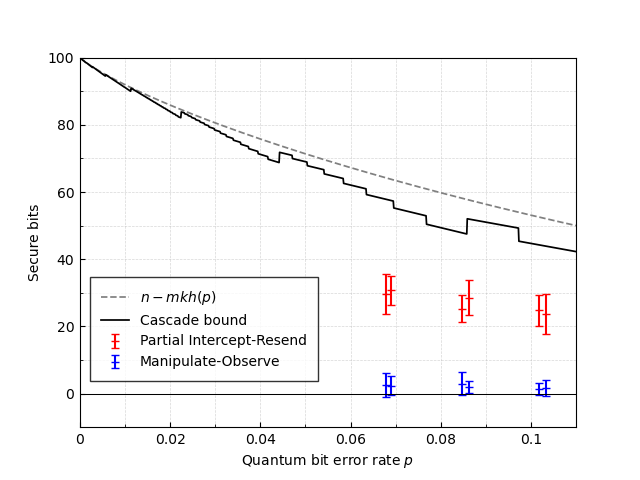}
        \subcaption{$\rho = 0.9$}
    \end{subfigure}
    \hfill
    \begin{subfigure}[b]{0.49\linewidth}
        \includegraphics[trim=20 8 40 25, clip, width=\linewidth]{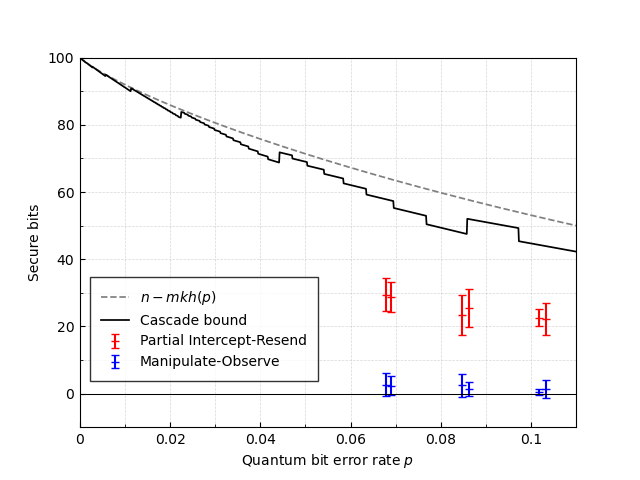}
        \subcaption{$\rho = 1.0$}
    \end{subfigure}
    \caption{The number of secure bits after reconciliation, using partial intercept-resend and \textbf{Manipulate-and-Observe Attacks} with different eavesdropping rates $\rho$, across various levels of estimated QBER $p$. The security is compared with Cascade's bound and the theoretical maximum security after reconciliation. Error bars intersecting zero secure bits indicate the complete unveiling of security, and a negative security value corresponds to an eavesdropper's redundant knowledge of the reconciled key material.}
    \label{Fig11}
\end{figure}

The performance of the Manipulate-and-Observe Attack is highlighted in Figure~\ref{Fig11}. Across all eavesdropping rates, the security after reconciliation is reduced to minuscule levels, significantly below Cascade's bound. Similarly, the attacks yielded many additional bits after the partial intercept-resend attacks, demonstrating the performance of the passive attack on Cascade. Notably, the full attacks obtained around 45-55 bits below the theoretical bound for Cascade. Moreover, the error bars cross zero for $\rho>0.7$, indicating that the security after reconciliation is entirely overcome. However, for $\rho=0.7$, the error bars cross zero for a single value of $p$. Despite that, the post-reconciliation security for $\rho=0.7$ is still minimal. 

However, it must be noted that hardware limitations constrain these results. Firstly, we could not obtain data for smaller values of QBER. The lack of data is apparent in the lower eavesdropping rates. Running the passive phase of the attack on reconciled keys with small values of QBER proved to be computationally infeasible for the available memory. The same was observed for values of QBER that were tested, where the maximum search space for the available memory was sometimes exceeded. These cases resulted in memory errors, and the lack of data could explain the constant level of security across all values of $\rho$ and $p$.

Another constraint on the results was the small pre-reconciled key size, which limited the number of Cascade passes to 3. However, the results show that three passes were sufficient to leak enough information to reduce the post-reconciliation security to minuscule levels. These results agree with those of \cite{mehicCalculationKeyLength2015, mehicErrorReconciliationQuantum2020}, where the final pass corrects negligible errors, leaking little information. Simulations of the attack for larger reconciled key sizes are required to investigate practical QKD systems.

Having said this, it is important to note that there are several countermeasures that can be taken to mitigate the effectiveness of the attack discussed here, namely:
\begin{itemize}
    \item Increasing the number of bits that are processed in the first round of Cascade would make it computationally infeasible to filter the reconciled keys during the passive phase.
    \item Increasing the sample size used in the QBER estimation step would provide more accurate QBER estimates, reducing the attack's success frequency.
    \item Reducing the QBER threshold to 6\% would thwart the attack in its current form using the same hardware and simulation parameters.
    \item One could monitor Cascade's interactivity metrics,  such as the number of calls to Binary or the look-back sub-protocol, and the distribution of parity mismatches. These metrics could be used to estimate the number of errors Cascade has corrected. If the newly estimated QBER anomalously exceeds BB84's QBER threshold, then the session could be aborted.
\end{itemize}

These modifications would have implications on the rate of secret key material that can be distilled and on the execution time of post-processing. Decreasing the QBER threshold would mean reducing the error tolerance of equipment during transmission, but the error rate of most current devices is already very low and therefore the threshold could be safely reduced. Note that we have only investigated one specific attack strategy, it is likely that there are others that could be applied. 

Finally, while reconciliation variants differ in implementation details, the structural leakage of parity constraints is common to all parity‑based schemes, and therefore the attack surface identified here is not specific to Cascade but inherent to this general class of protocols.

\section{Conclusions}
In this paper, we have presented a new attack that we are calling the 'Manipulate-and-Observe Attack' applied to the combination of the quantum, BB84, protocol and the classical exchange, Cascade. In this attack a partial intercept-resend eavesdropping strategy is used during the quantum key exchange to obtain information that we can exploit during post-processing. The aim of the partial eavesdropping strategy is to inject the maximum tolerable error rate into the raw key material that will become the secret key. This ensures that Cascade leaks the maximum amount of information, while the attacker remains undetected. A passive attack filtered a set of possible reconciled key candidates by exploiting Cascade's leaked parity information. Our simulation results demonstrate that the full attack can completely reduce the security of the reconciled key. Any remaining security was negligible and significantly lower than Cascade's theoretical bound for all accessible values of QBER \cite{brassardSecretKeyReconciliationPublic1994}.

This investigation necessarily operated within computational limits imposed by the processor and memory that was used to perform the attack. Most notably this limited us to the use of 100‑bit reconciled keys, a reduced number of Cascade passes and a bounded search-space exploration. However, these constraints reflect the practical realities of demonstrating a proof‑of‑concept attack whose complexity inherently scales with parity leakage and block structure. Crucially, the success of the attack in these constrained scenarios highlights a structural vulnerability in the extensible security assumptions of BB84 and Cascade, independent of scale. If anything, larger keys, more varied pass structures and higher‑capacity compute architectures (e.g., GPU clusters or distributed filtering systems) allow for a more complete exploration of the attack surface and may amplify — rather than diminish — the observed effects. We therefore view this result as a natural stepping‑stone toward a full characterisation of the composability gap between quantum transmission and classical reconciliation in QKD.

Our results show that there exist edge cases in which the combined security of BB84 and Cascade can be entirely overcome by the attack. Our work presents a proof-of-concept of the attack, which could be scaled up with appropriate hardware. By implementing the attack with a low-level language like C++, the filtering process could be sped up by orders of magnitude. Since the parity checks use XOR operations, a GPU may be suited to run some parts of the simulations. Further parallelisation of the filtering process across a GPU cluster with access to significant memory could yield massive performance improvements and overcome the security limitations of larger reconciled keys. The attack may also be effective against other reconciliation protocols that compare subset parities, like LDPC. Therefore, with adequate computational resources, the Manipulate-and-Observe Attack could pose a serious threat to QKD systems, and an integrated threat model of BB84 and Cascade is urgently recommended.

\bibliographystyle{apsrev4-2}

\titleformat{\Section}[display]
  {\bfseries\Huge}
  {\ifnum\value{chapter}>0 Appendix \thechapter \else \thechapter \fi}
  {1ex}
  {\Huge}
\begin{appendices}
\section{Cascade}
For calibration purposes we set out the version of cascade that is being used in this paper. Cascade succeeds a former reconciliation protocol Binary, sometimes referred to as BBBSS, presented in \cite{bennettExperimentalQuantumCryptography1992}. Binary is a highly interactive binary search algorithm fundamental to Cascade \cite{brassardSecretKeyReconciliationPublic1994}. The protocol is applied to blocks of bits containing an odd number of errors, which are reconciled using parity checks \cite{brassardSecretKeyReconciliationPublic1994}. Its procedural steps are as follows: \\

\noindent\textbf{A. \quad Binary} 
\begin{enumerate}[leftmargin=*]
    \item Alice transmits the parity of the first half of her block to Bob \cite{bennettExperimentalQuantumCryptography1992}.
    \item Bob compares the parity of the first half of his block with Alice's to determine whether the first or second half of his block contains an error \cite{bennettExperimentalQuantumCryptography1992}.
    \begin{enumerate}
        \item If the parities match, the error is in the second half of the block.
        \item Else, the error is in the first half of the block.
    \end{enumerate}
    \item Steps 1 and 2 are repeated on the half that was determined to contain the error until an error is located \cite{bennettExperimentalQuantumCryptography1992}.
    \item The error is corrected by flipping the bit value \cite{brassardSecretKeyReconciliationPublic1994}.
\end{enumerate}

Note that if the block size is odd, the left sub-block is 1 bit greater than the right sub-block \cite{PDFCrackingCurious2024}. Additionally, the parities are exchanged via parity bits; the final bit in each block/sub-block \cite{mehicErrorReconciliationQuantum2020}.

Cascade aims to distribute errors throughout the sifted key evenly and split it into blocks, where Alice and Bob compare the parities of their respective blocks via the classical channel \cite{brassardSecretKeyReconciliationPublic1994}. If the parities of a block mismatch, Bob's block must contain an odd number of errors, and Binary is run \cite{brassardSecretKeyReconciliationPublic1994}. The protocol performs a preset number of passes, and the steps for pass 1 are as follows:\\

\noindent\textbf{B. \quad Cascade}
\begin{enumerate}[leftmargin=*]
    \item Alice and Bob divide their sifted keys into blocks of constant size $k_1$, given by $k_1 = 0.73/p$ \cite{mehicErrorReconciliationQuantum2020}.
    \item Alice and Bob compute the parities of their top blocks from step 1 \cite{brassardSecretKeyReconciliationPublic1994}.
    \item The parities of the blocks are publicly compared \cite{brassardSecretKeyReconciliationPublic1994}.
    \item For any parity mismatches, Binary is run on Bob's respective block to locate and correct an error \cite{brassardSecretKeyReconciliationPublic1994}.
\end{enumerate}

Note that any remaining bits after the sifted key is split form a remainder block, which is dealt with similarly to the other top blocks \cite{PDFCrackingCurious2024}. After the first pass, Bob's top blocks will all contain an even number of errors, possibly zero \cite{brassardSecretKeyReconciliationPublic1994}. If a block contains an even number of errors, then they will be undetected by the parity checks. Thus, for passes $i>1$, the bits are randomly permuted to disperse any errors evenly \cite{mehicErrorReconciliationQuantum2020}. Moreover, the block size doubles from the previous pass: $k_i = 2  k_{i-1}$ \cite{mehicErrorReconciliationQuantum2020}. These passes continue with steps 2 and 3 from the first pass, where the bits are partitioned into blocks, and the parities of the blocks are compared \cite{brassardSecretKeyReconciliationPublic1994}. 

However, whenever an error is reconciled with Binary, a paired error can be dealt with using a recursive look-back sub-protocol \cite{pedersenHighPerformanceInformation2013}. For any error corrected in pass $i>1$, there must be, at minimum, one other error positioned in the same blocks in all previous passes because they were masked in the previous parity checks \cite{mehicCalculationKeyLength2015}. Thus, for any error corrected in passes $i>1$, the corresponding blocks in all previous passes containing that error are added to a look-back list \cite{pedersenHighPerformanceInformation2013}. Binary is applied to each block in the list with a parity mismatch, starting with the smallest block from pass 1 \cite{pedersenHighPerformanceInformation2013}. For each new error corrected during the look-back sub-protocol, all the blocks from all previous passes containing the new error are also added to the look-back list \cite{pedersenHighPerformanceInformation2013}. This sub-protocol terminates when the list is empty, and Cascade continues with the following top block in the current pass \cite{pedersenHighPerformanceInformation2013}. The discovery of new errors during look-back sparks a chain reaction that cascades through all the previous passes, hence the protocol's name.

Four passes are used in the original version of Cascade \cite{brassardSecretKeyReconciliationPublic1994}. However, it is suggested that the block sizes should increase until they are half the sifted key size \cite{mehicCalculationKeyLength2015}. Furthermore, various modifications have been developed that have altered the initial block size, scaling of blocks in subsequent passes, number of passes, and other optimisations like the reuse of blocks \cite{martinez-mateoDemystifyingInformationReconciliation2014}.

Cascade is sometimes criticised because redundant information is exchanged during each pass, which increases the protocol's computational demand \cite{buttlerFastEfficientError2003}. Other protocols like Winnow use privacy maintenance where some bits are discarded during reconciliation \cite{buttlerFastEfficientError2003}. However, keeping the redundant information enables more errors to be corrected via the look-back sub-protocol \cite{brassardSecretKeyReconciliationPublic1994}. More recent considerations of this issue advise discarding the parity bits after execution of the protocol \cite{mehicErrorReconciliationQuantum2020}.

\section{Passive Attack Simulation Results}
Cascade's random bit shuffling for passes $i>1$ can be shown to vary an eavesdropper's search space during the passive attack outlined in Section 2.2. Table~\ref{Table 5} provides simulation parameters for Cascade, where six errors have been inserted into Bob's pre-reconciled key. We impose an initial block size of 4 bits, where the first three blocks contain a single error that will be corrected during the first pass. The final block contains three errors placed next to each other. One of these errors will be corrected during the first pass, masking the other two in the process. During pass 2, the masked errors will be corrected via the look-back sub-protocol.

\begin{table}[h]
    \centering
    \begin{tabular}{l l}
        \hline
        \textbf{Property} & \textbf{Value} \\
        \hline
        Alice's initial key & 1111000011110000 \\
        Bob's initial key & 1110000111100111 \\
        Bit IDs & 0-15 \\
        Cascade passes & $2$ \\
        Initial block size & $4$ bits \\
        \hline
    \end{tabular}
    \caption{\label{Table 5}Simulation parameters for Cascade.}
\end{table}

Tables~\ref{Table 6}--\ref{Table 9} show that the different permutations during the second pass of Cascade can lead to various search spaces for an eavesdropper. Table 6 demonstrates that some permutations can lead to the complete reconstruction of the reconciled key. The results indicate that the random permutations give an extra constraint during filtering.

\begin{table}[h]
    \centering
    \begin{tabular}{l l}
        \hline
        \textbf{Property} & \textbf{Value} \\
        \hline
        Bob's reconciled key & 1111000011110000 \\
        Permuted bit IDs & [15, 1, 5, 13, 8, 11, 0, 2], [7, 6, 4, 9, 14, 12, 10, 3] \\
        Leaked information & 19 bits \\
        Leaked parity bit IDs & 1, 2, 3, 4, 5, 6, 7, 9, 10, 11, 12, 13, 14, 15 \\
        Eve's valid keys & 1111000011110000 \\
        Eve's search space & 1 valid key \\
        \hline
    \end{tabular}
    \caption{\label{Table 6}Simulation results of the passive attack on Cascade, where the random permutation results in a search space of 1 valid key.}
\end{table}

\begin{table}[h]
    \centering
    \begin{tabular}{l l}
        \hline
        \textbf{Property} & \textbf{Value} \\
        \hline
        Bob's reconciled key & 1111000011110000 \\
        Permuted bit IDs & [1, 6, 8, 9, 13, 4, 2, 14], [10, 7, 15, 11, 3, 0, 5, 12] \\
        Leaked information & 19 bits \\
        Leaked parity bit IDs & 1, 2, 3, 5, 6, 7, 9, 10, 11, 12, 13, 14, 15 \\
        Eve's valid keys & 1111000000110000, 1111000011110000 \\
        Eve's search space & 2 valid keys \\
        \hline
    \end{tabular}
    \caption{\label{Table 7}Simulation results of the passive attack on Cascade, where the random permutation results in a search space of 2 valid keys.}
\end{table}

\begin{table}[h]
    \centering
    \begin{tabular}{l l}
        \hline
        \textbf{Property} & \textbf{Value} \\
        \hline
        Bob's reconciled key & 1111000011110000 \\
        Permuted bit IDs & [7, 13, 4, 1, 6, 2, 12, 15], [5, 0, 11, 14, 3, 9, 8, 10] \\
        Leaked information & 19 bits \\
        Leaked parity bit IDs & 1, 2, 3, 5, 6, 7, 9, 10, 11, 12, 13, 14, 15 \\
        Eve's valid keys & 0011110000110000, 0011110011110000, \\
         & 1111000000110000, 1111000011110000 \\
        Eve's search space & 4 valid keys \\
        \hline
    \end{tabular}
    \caption{\label{Table 8}Simulation results of the passive attack on Cascade, where the random permutation results in a search space of 4 valid keys.}
\end{table}

\begin{table}[h]
    \centering
    \begin{tabular}{l l}
        \hline
        \textbf{Property} & \textbf{Value} \\
        \hline
        Bob's reconciled key & 1111000011110011 \\
        Permuted bit IDs & [3, 13, 7, 2, 6, 10, 4, 1], [14, 0, 15, 9, 8, 12, 11, 5] \\
        Leaked information & 14 bits \\
        Leaked parity bit IDs & 1, 2, 3, 5, 6, 7, 9, 10, 11, 12, 13, 15 \\
        Eve's valid keys & 0011110000110000, 0011110000110011, \\
         & 0011110011110000, 0011110011110011, \\
         & 1111000000110000, 1111000000110011, \\
         & 1111000011110000, 1111000011110011 \\        
        Eve's search space & 8 valid keys \\
        \hline
    \end{tabular}
    \caption{\label{Table 9}Simulation results of the passive attack on Cascade, where the random permutation results in a search space of 8 valid keys. Note that the last two errors were masked during the second pass and were not corrected.}
\end{table}

\clearpage
\section{Optimised Key Filtering}
We can optimise the brute force method by searching an array of blocks where the parity conditions match. Since $k$ unique parity checks are required to reveal a unique $k$ bit block, we can encode every possible block with its corresponding unique parity checks. We demonstrate such an encoding for 4-bit blocks in Table~\ref{Table 1}. The collection of the unique parity checks forms a secondary block, referred to as the parity block. The parity bit indices for each unique parity check order such a block.

\renewcommand{\arraystretch}{1.3}
\setlength{\tabcolsep}{5pt}

\begin{table}[h]
    \centering
    \begin{tabular}{|>{\centering\arraybackslash}p{2.25cm}|>{\centering\arraybackslash}p{0.75cm}|>{\centering\arraybackslash}p{0.75cm}|>{\centering\arraybackslash}p{0.75cm}|>{\centering\arraybackslash}p{0.75cm}|>{\centering\arraybackslash}p{2.25cm}|}
        \hline
        \textbf{Block} & \(P_0\) & \(P_{01}\) & \(P_2\) & \(P_{0123}\) & \textbf{Parity Block} \\
        \hline
        {[0 0 0 0]} & 0 & 0 & 0 & 0 & {[0 0 0 0]} \\
        {[0 0 0 1]} & 0 & 0 & 0 & 1 & {[0 0 0 1]} \\
        {[0 0 1 0]} & 0 & 0 & 1 & 1 & {[0 0 1 1]} \\
        {[0 0 1 1]} & 0 & 0 & 1 & 0 & {[0 0 1 0]} \\
        {[0 1 0 0]} & 0 & 1 & 0 & 1 & {[0 1 0 1]} \\
        {[0 1 0 1]} & 0 & 1 & 0 & 0 & {[0 1 0 0]} \\
        {[0 1 1 0]} & 0 & 1 & 1 & 0 & {[0 1 1 0]} \\
        {[0 1 1 1]} & 0 & 1 & 1 & 1 & {[0 1 1 1]} \\
        {[1 0 0 0]} & 1 & 1 & 0 & 1 & {[1 1 0 1]} \\
        {[1 0 0 1]} & 1 & 1 & 0 & 0 & {[1 1 0 0]} \\
        {[1 0 1 0]} & 1 & 1 & 1 & 0 & {[1 1 1 0]} \\
        {[1 0 1 1]} & 1 & 1 & 1 & 1 & {[1 1 1 1]} \\
        {[1 1 0 0]} & 1 & 0 & 0 & 0 & {[1 0 0 0]} \\
        {[1 1 0 1]} & 1 & 0 & 0 & 1 & {[1 0 0 1]} \\
        {[1 1 1 0]} & 1 & 0 & 1 & 1 & {[1 0 1 1]} \\
        {[1 1 1 1]} & 1 & 0 & 1 & 0 & {[1 0 1 0]} \\
        \hline
    \end{tabular}
    \caption{\label{Table 1}Encoding all 4-bit blocks using all the unique parity checks that Cascade and Binary can perform.}
\end{table}

The block search approach can be vectorised to enable Eve to check all parity conditions simultaneously rather than filtering the blocks one condition at a time. Furthermore, the array of possible blocks can be filtered similarly for any bits revealed or obtained earlier in the QKD protocol. Hence, the vectorised block search approach is compatible with active eavesdropping strategies on the quantum phase of QKD. 

We apply the vectorised block search method to filter the blocks in the first pass of Cascade because the first pass typically leaks the most information, as it uses the smallest block size and corrects the greatest number of errors. After the search, we are left with a set of valid blocks that satisfy the parity conditions for Alice's corresponding blocks. The valid blocks in each set are combined to generate the valid keys after pass 1. The valid keys after pass 1 are filtered using the remaining parity constraints from the additional passes.

We can make further optimisations using parallelisation. Firstly, the blocks in pass 1 can be searched for in parallel. Secondly, the array of valid keys after pass 1 can be split into chunks to be processed in parallel. After filtering the chunks, the remaining pass 1 valid keys are recombined to obtain the final array of valid keys that satisfy all the leaked parity conditions.

Detailed steps of the passive phase are as follows:
\begin{enumerate}[leftmargin=*]
    \item Eve stores the leaked information for each parity comparison made in Cascade/Binary, including the pass index, the top block index, the indices of the bits and the parity.
    \item Simultaneously, Eve stores an ID for each bit in Alice's pre-reconciled key, to keep track of the location of each bit after shuffling in each pass.
    \item When Cascade ends, Eve determines the initial block size $k_1$ by looking at the size of the first top block in pass 1 and assesses whether a remainder block is required.
    \item Eve divides her initial key into blocks of size $k_1$ (if no eavesdropping is done before the passive attack, then she will have no bits at this stage).
    \item Eve loads an array that contains all the possible $k_1$ bit blocks and their corresponding parity blocks.
    \item In parallel, an array of valid blocks is determined for each block index in pass 1 using the vectorised block search:
    \begin{enumerate}[label=\alph*]
        \item The indices and corresponding bit values in Eve's initial block are determined.
        \item The array of possible blocks and secondary parity blocks are filtered by Eve's initial bits.
        \item The array of secondary parity blocks is filtered using the parity data leaked by Cascade, and the corresponding primary blocks are taken as valid blocks.
    \end{enumerate}
    \item Each array of valid blocks is combined to form an array of valid keys for pass 1, and the number of valid keys is counted to obtain the pass 1 search space.
    \item The array of valid keys is split into chunks to be processed in parallel.
    \item Brute force is used to filter the pass 1 valid keys in each chunk:
    \begin{enumerate}[label=\alph*]
        \item For each Cascade pass $i>1$ and for each key in the chunk:
        \begin{enumerate}[label=\roman*]
            \item The blocks for the current key are constructed using the shuffled bit IDs.
            \item The blocks are validated by applying the leaked parity conditions.
            \item The key is valid if all its blocks are valid.
        \end{enumerate}
        \item The number of valid keys for the current pass is counted.
    \end{enumerate}
    \item The valid keys in each processed chunk are recombined, and the number of valid keys after each pass is computed to get the corresponding search space.
\end{enumerate}

\section{Additional Details of Results}
\subsection{Partial-Intercept Resend}
\begin{figure}[H]
\centering
\includegraphics[trim=0 0 0 25, clip, width=0.75\linewidth]{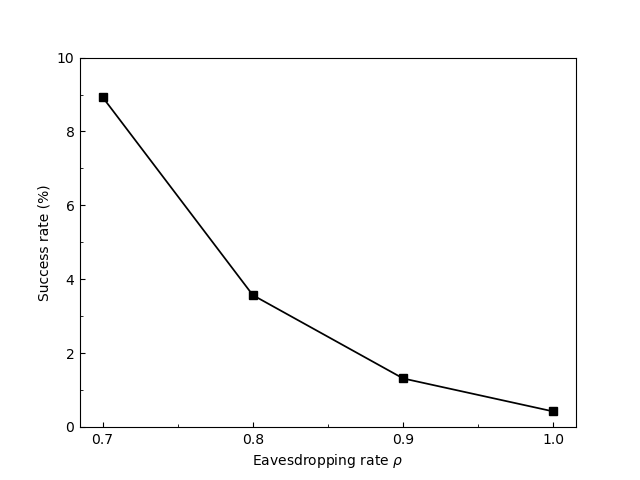}
\caption{\label{Fig9} The success rate of the partial intercept-resend attack for various eavesdropping rates. An attack is successful if the eavesdropper is not detected during QBER estimation. Thus, the estimated QBER must be less than 0.11.}
\end{figure}

Figure~\ref{Fig9} shows the success rates for the partial intercept-resend attacks on the quantum phase of BB84 for different eavesdropping rates $\rho$. The figure demonstrates that the success rate decreases as the eavesdropping rate increases. As $\rho$ approaches $1$, the success rate approaches zero. However, for larger $n$, the success rate is expected to be significantly lower across all levels of $\rho$, as better statistics will be obtained from a larger sample size whilst estimating the QBER.

\begin{table}[h]
    \centering
    \begin{tabular}{|c|c|c|c|c|c|c|}
        \hline
        \diagbox[width=6cm]{\textbf{Eavesdropping rate}}{\textbf{QBER}} & \textbf{0.068}  & \textbf{0.069} & \textbf{0.085} & \textbf{0.086} & \textbf{0.102} & \textbf{0.103} \\
        \hline
        \textbf{0.7} & & & 28.84 & 32.31 & 27.67 & 28.53 \\
        \hline
        \textbf{0.8} & & 33.64 & 27.73 & 31.09 & 26.12 & 26.52 \\
        \hline
        \textbf{0.9} & 29.68 & 30.71 & 25.29 & 28.55 & 24.77 & 23.71 \\
        \hline
        \textbf{1.0} & 29.44 & 28.79 & 23.33 & 25.51 & 22.63 & 22.26 \\
        \hline
    \end{tabular}
    \caption{\label{Table 3}The number of secure bits after reconciliation, using partial intercept-resend attacks for different eavesdropping rates $\rho$, across various levels of estimated QBER $p$. Results cover the levels of QBER and eavesdropping relevant to the Manipulate-and-Observe Attack.}
\end{table}

Table~\ref{Table 3} shows that the security after reconciliation decreases linearly as the eavesdropping rate increases. For values of QBER above 0.08, there are around 6 less secure bits after reconciliation for full eavesdropping than at 70\%. Similarly, the security tends to decrease as the QBER increases. 

\subsection{Manipulate-and-Observe}
\begin{table}[h]
    \centering
    \begin{tabular}{|c|c|c|c|c|c|c|}
        \hline
        \diagbox[width=6cm]{\textbf{Eavesdropping rate}}{\textbf{QBER}} & \textbf{0.068}  & \textbf{0.069} & \textbf{0.085} & \textbf{0.086} & \textbf{0.102} & \textbf{0.103} \\
        \hline
        \textbf{0.7} & & & 2.82E-04 & 3.44E-04 & 5.96E-04 & 7.72E-04 \\
        \hline
        \textbf{0.8} & & 4.35E-05 & 2.03E-04 & 3.18E-04 & 4.47E-04 & 1.24E-03 \\
        \hline
        \textbf{0.9} & 2.16E-05 & 3.21E-05 & 9.05E-05 & 1.21E-04 & 2.74E-04 & 4.15E-04 \\
        \hline
        \textbf{1.0} & 7.72E-06 & 8.73E-06 & 2.41E-05 & 5.71E-05 & 1.09E-04 & 1.59E-04 \\
        \hline
    \end{tabular}
    \caption{\label{M-O SR}The success rate of the Manipulate-and-Observe attack for various eavesdropping rates and QBER. An attack is successful if the partial intercept-resend attack is undetected and enough computational resources, like memory, are available to run the passive phase of the attack. The success rate is determined by dividing the number of simulation runs (20) by the final seed used.}
\end{table}

Table~\ref{M-O SR} shows the success rates for the Manipulate-and-Observe attacks for different eavesdropping rates $\rho$ and levels of QBER $p$. The low success rates do not mean that BB84 and Cascade are completely secure against the attack, as we demonstrate that edge cases exist in which we can entirely undermine their security. The lowest success probability is 0.000772\%, implying that, on average, the attack would succeed roughly once every 130,000 attempts. This frequency is more significant when contextualised, as millions of digital communications are transmitted each day. Moreover, the success frequency is expected to increase if greater computational resources are available.

\begin{table}[h]
    \centering
    \begin{tabular}{|c|c|c|c|c|c|c|}
        \hline
        \diagbox[width=6cm]{\textbf{Eavesdropping rate}}{\textbf{QBER}} & \textbf{0.068}  & \textbf{0.069} & \textbf{0.085} & \textbf{0.086} & \textbf{0.102} & \textbf{0.103} \\
        \hline
        \textbf{0.7} & & & 2.65 & 3.35 & 2.70 & 2.95 \\
        \hline
        \textbf{0.8} & & 1.75 & 1.10 & 1.70 & 1.90 & 2.55 \\
        \hline
        \textbf{0.9} & 2.45 & 2.35 & 2.95 & 1.85 & 1.35 & 1.70 \\
        \hline
        \textbf{1.0} & 2.70 & 2.35 & 2.45 & 1.50 & 0.55 & 1.35 \\
        \hline
    \end{tabular}
    \caption{\label{Table 4}The number of secure bits after reconciliation, using Manipulate-and-Observe Attacks for different eavesdropping rates $\rho$, across various levels of estimated QBER $p$. The missing values are where the attacks failed to obtain 20 results due to hardware limitations.}
\end{table}

Table~\ref{Table 4} presents the number of secure bits after reconciliation after deploying the Manipulate-and-Observe Attack for various $\rho$ and $p$. The security is minimal for all values of $\rho$ and $p$. Furthermore, the security is significantly lower than that obtained by the partial-intercept resend attacks. However, the security appears constant across the table with no clear relationship with either $\rho$ or $p$.

\end{appendices}

\end{document}